\documentclass[12pt]{article}
\usepackage{a4,amsmath,epsfig,amssymb}
\usepackage[latin1]{inputenc}
\usepackage{graphicx}
\usepackage{caption}
\usepackage{subcaption}
\usepackage{feynmp}
\usepackage{amsmath}
\begin{document}
\bigskip\bigskip

\begin{center}
\begin{large}
{\large \bf Polarization in (Quasi-)Two-Body Decays\\
and New Physics}
\end{large}
\end{center}

\vspace{8pt}
\begin{center}
\begin{large}
  
Z.~J.~Ajaltouni$^{a,}$\footnote{ziad@clermont.in2p3.fr},
E.~Di Salvo$^{a,b,}$\footnote{Elvio.Disalvo@ge.infn.it}
\end{large} 

\bigskip

$^a$ 
Laboratoire de Physique Corpusculaire de Clermont-Ferrand, \\
IN2P3/CNRS Universit\'e Blaise Pascal, 
F-63177 Aubi\`ere Cedex, France\\

\noindent  
$^b$ 
Dipartimento di Fisica dell' Universita' di Genova \\
and I.N.F.N. - Sez. Genova,\\
Via Dodecaneso, 33, 16146 Genova, Italy \\  

\noindent  

\vskip 1 true cm
\end{center}
\vspace{1.0cm}

\vspace{6pt}
\begin{center}{\large \bf Abstract}

We consider two-body and quasi-two-body decays of the type $f_1 \to f_2  B$, where $f_1$ and $f_2$ are spin-1/2 fermions and $B$ a spin-0 or spin-1 boson. After recalling the non-covariant formalism for decay amplitudes, we derive the expression of the differential decay width and of the polarizations of the final spinning particles, both on- and off-shell. We find an intriguing geometrical interpretation of the results about the polarization. We also illustrate some methods for measuring the polarizations of the resonances and for optimizing data analysis. Then we propose applications to semi-leptonic weak decays, with a major attention to the $T$-odd component of the polarization; this may help to find, simultaneously, possible time-reversal violations and hints to physics beyond the standard model. We suggest also a $CPT$ test. Last, we discuss some $T$-odd observables for the production process of $f_1$ and for the study of the strong final state interactions of non-leptonic decays.
 
\end{center}

\vspace{10pt}

\centerline{PACS numbers: PACS Nos.: 11.30.Er, 11.80.Et, 13.25.-k, 13.30.Eg}

\newpage

\section{Introduction}
The search for new physics (NP) beyond the standard model (SM) is especially stimulating now, that the Higgs boson has been found\cite{cms,atl}. Indeed, the SM has achieved a resounding success, thanks to the wealthy of experimental confirmations. However, it presents several unsatisfactory aspects[3-5] and has to be regarded, at best, as an effective low-energy approximation\cite{alt} of the theory of the fundamental particles. Therefore it is essential to realize experiments where, either a clear contradiction with the SM is found, or at least more stringent constraints on physics beyond it are established. Indeed, in the past few years, some data concerning semi-leptonic decays exhibited strong tensions with the SM predictions[7-11].

Confirmations to such hints are expected from the polarization of spinning particles produced in weak decays, particularly the semi-leptonic ones, like
\begin{equation}
t \to b \ell \nu_{\ell}, ~~~ ~~~ \Lambda_b \to \Lambda_c \ell \bar{\nu_{\ell}}.   \label{sld}
\end{equation}
Here $\ell$ denotes a charged lepton, including $\tau$, and $\nu_{\ell}$ ($\bar{\nu_{\ell}}$) the corresponding (anti-)neutrino. Parity violation in the production of the decaying particles favors their polarization: the
$\Lambda_b$ (and $\bar{\Lambda}_b$) coming from the transition $e^+ e^- \to Z \to b\bar{b}$ \cite{op,de} have a
sizable polarization and the top produced in the $t$-channel is strongly polarized\cite{atl2}. 

These circumstances offer the opportunity of defining new observables which are sensitive to NP. For example, in the $\Lambda_b$ decay, the polarization of the charged lepton or of the final baryon may reveal deviations from the SM\cite{gu1}. However, given the wealth of available data about the top quark, and its presumable coupling to new particles, the high energy physicists are mainly attracted by the mechanisms of production and decay of this particle. Indeed, the importance of such processes was stressed even before its very discovery[16-19]. Successively, a lot of papers were dedicated both to the decay[20-26] and to the production[27-31] of the top quark, either performing precise SM predictions with QCD corrections\cite{fi1,fi2,gr1,gr2,kal}, or suggesting tests for physics beyond the SM\cite{sh,cho,cao,ab1,ab2,age}.

An especial attention deserve the $T$-odd\footnote{We refer, throughout the paper, to the {\it motion reversal} operator\cite{dug}, which includes both real time reversal and certain rotations that invert momenta and spins\cite{der}} observables, like the normal component of the polarization\cite{ajd}, for which, in a semi-leptonic decay of the type (\ref{sld}), the SM predicts quite a small value\cite{fi3}. Indeed, in this model, decays like (\ref{sld}) proceed only through the intermediate vector boson $W$, therefore no time reversal violation (TRV) occurs. However, tiny (and calculable\cite{jk1,iv2}) $T$-odd effects are produced by final-state electromagnetic spin-orbit\cite{br1,br2} interactions, giving rise to a negligibly small $T$-odd ({\it naive} $T$\cite{dug}) polarization. Therefore, a sizable value of such an observable would be a simultaneous signature of NP\cite{ab1,ab2,ka,jek,ag3} and of TRV\cite{ga}. This may happen if some non-standard coupling[16,18,23,42-46], like a charged higgs exchange[47-52] or $V+A$\cite{ab2,jek}, has a non-trivial relative phase with respect to the SM amplitude. 

In the present paper, we examine in detail the polarizations of the initial and final particles in a (quasi-)two-body decay of the type
\begin{equation}
f_1 \to f_2  B,   \label{ffv}
\end{equation}
where $f_i$ are spin-1/2 fermions and $B$ is a spin-0 ($S$) or spin-1 ($V$) boson. This includes the decays (\ref{sld}) - assuming they occur through the $W$-boson and through the charged Higgs $H$ -, and non-leptonic decays like
\begin{equation}~~~
\Lambda_b \to \Lambda J/\psi, ~~~ \Lambda_b \to \Lambda_c \rho^-, ~~~ \Lambda_c \to \Lambda \pi^+(K^+). \label{nld}
\end{equation}

Our approach is based on the non-covariant formalism of the spin density matrix, analogously to Aguilar and Bernabeu\cite{ab1,ab2}. However, we propose a more general method, applicable to the decays (\ref{sld}) and (\ref{nld}), which involve both on-shell and off-shell bosons; moreover, in an optics of NP contribution to semi-leptonic decays, we take into account the interference between a vector and a scalar boson. Last, we find some intriguing features of the polarizations of the final particles. In particular, we establish a relation between some measurable angle and a given quantum-mechanical phase; moreover, we single out some $T$-odd observables, including the above mentioned angle, that are sensitive to NP; we also indicate the most favorable conditions for exploiting the available data, even when the statistics is not so rich.   

Our main aim is to apply such results to specific semi-leptonic decays of the type (\ref{ffv}), quite suitable for testing the SM; however, also non-leptonic decays like (\ref{nld}) could give important information about the $T$-odd effects of the strong interactions. Moreover, as we shall see, also the polarization of the parent resonance may help to elaborate tests of NP. 


We dedicate Sect. 2 to the non-covariant formalism for the decay amplitudes, considering both on- and off-shell bosons. In Sects. 3 and 4, we deduce the expressions of the differential decay widths and of the polarizations for the decays (\ref{ffv}), with on-shell scalar and vector bosons respectively. In Sect. 5, we generalize our results to the case of off-shell bosons, considering also the possibility of an interference between $V$ and $S$. In Sect. 6, we show some methods for measuring the polarizations of the final particles. Sect. 7 is devoted to illustrating the optimization of the data analysis. In Sect. 8, we discuss some applications to the search for NP and to the experimental determination of $T$-odd strong interactions. Last, we draw some conclusions in Sect. 9.


\section{Decay Amplitudes}

\subsection{Formalism}

The non-covariant decay amplitudes provide a useful parametrization for the experimental determination of the differential decay widths of two- and three-body decays\cite{jw}. Of course, a correct description of such processes requires Lorentz Invariance (LI) of the amplitudes\cite{mi}. However, in practice, these are constructed from four-vectors, which must be defined in a given frame; this last, if accurately chosen, makes the parametrization of the observables especially simple and intuitive.

In the framework of the  non-covariant amplitudes, the helicity representation presents some advantages over the $L-S$ one\cite{jw}:

- one avoids the complication of splitting the angular momentum into an orbital and a spin component;

- the helicities of the final particles are rotationally invariant, which allows to construct states with a definite angular momentum, while preserving the individual polarization properties of each particle;

- the decay amplitudes are themselves rotationally invariant.

The non-covariant helicity amplitudes were connected in the past years to LI form factors both for semi-leptonic and non-leptonic two-body decays of the type (\ref{ffv})[55-59], with the aim of introducing some models, based on some significant Feynman diagrams. Here we propose a slightly different version of such a treatment, which leads to some results, to be used in the successive sections. We consider the general case, where the boson $B$ is not necessarily on-shell and decays to more stable particles.
  
\subsection{$f_1$ $\to$ $f_2$ $S$ Decay Amplitude}

We consider, not only the two-body decay of $f_1$, whose amplitude we denote by $h_S$, but also some successive decay mode of $S$, described by the amplitude $\ell_S$. Then the overall amplitude of the process reads as
\begin{equation}
A_S = h_S D \ell_S. \label{asd}
\end{equation}
Here $D$ is the propagator of $S$, {\it i. e.},
\begin{equation}
D = (Q^2-M^2+iM\Gamma)^{-1}, \label{prp}
\end{equation}
$M$ and $\Gamma$ being respectively the rest mass and the width of $S$ and $Q^2$ the modulus squared of the four-momentum of $S$. $h_S$ is parametrized as
\begin{equation}
h_S = {\bar u}_2 O_S u_1, \ ~~~~~~ \  O_S = {\cal F}_0(Q^2)+\gamma_5{\cal G}_0(Q^2), \label{haS}
\end{equation}
where $u_i$ is the Dirac four-spinor of the fermion $f_i$ ($i$ = 1, 2) and ${\cal F}_0$ and ${\cal G}_0$ are suitable form factors. Fixing a coordinate system in a frame at rest with respect to $f_1$, one has
\begin{equation}
h_S = {\hat h}_S(m;\lambda;{\vec p}) = \frac{1}{\sqrt{2\pi}} a_{\lambda}(p) 
{\cal D}^{1/2*}_{m-\lambda}(\phi,\theta,0). \label{ria}
\end{equation}
Here
\begin{equation}
\vec{p} \equiv (p ~ sin \theta ~ cos \phi, ~~ p ~ sin \theta ~ sin \phi, ~~ p ~ cos \theta) \label{mmt}
\end{equation}
is the momentum of the boson in the frame that we have fixed; moreover $m$= $\pm1/2$  and $\lambda$ = $\pm1/2$ are, respectively, the $z$-component of the spin of $f_1$ and the helicity of $f_2$. ${\cal D}^{1/2}_{m-\lambda}$ is an element of the Wigner rotation matrix for spin-1/2 objects. Last, $a_{\lambda}(p)$ is the rotationally invariant amplitude:
\begin{equation}
\frac{1}{\sqrt{2\pi}} a_{\lambda}(p) = F[{\cal F}_0(Q^2)+2\lambda\chi{\cal G}_0(Q^2)], ~~~~
F = \sqrt{2Q_1} \sqrt{E_2+Q_2},  ~~~~  \chi = \frac{p}{E_2+Q_2},
\end{equation}
$Q_i$ being the modulus of the four-momentum of $f_i$ ($i$ = 1,2) and $E_2$ = $\sqrt{p^2+Q_2^2}$.

\subsection{$f_1$ $\to$ $f_2$ $V$ Decay Amplitude}

We have, analogously to the previous case,
\begin{equation}
A_V = h_V^{\alpha} D_{\alpha\beta} \ell_V^{\beta}, \ ~~~~~~ \ D_{\alpha\beta} = D (k_{\alpha} k_{\beta}-g_{\alpha\beta}), \ ~~~~~~ \ k = \frac{q}{M}; \label{avd}
\end{equation}
here $q$ is the four-momentum of $V$, while the other symbols are analogous to those introduced above. $h_V^{\alpha}$ is parametrized as
\begin{eqnarray}
h_V^{\alpha} &=& {\bar u}_2 O_V^{\alpha} u_1, \ ~~~~~~ \  O_V^{\alpha} = \Gamma^{\alpha}+\Gamma^{\alpha}_5, \label{haV}
\\
\Gamma^{\alpha} &=& {\cal F}_1(Q^2)\gamma^{\alpha}+{\cal F}_2(Q^2)i\sigma^{\alpha\beta}q_{\beta} +{\cal F}_3(Q^2)q^{\alpha},
\\
\Gamma^{\alpha}_5 &=& [{\cal G}_1(Q^2)\gamma^{\alpha}+{\cal G}_2(Q^2)i\sigma^{\alpha\beta}q_{\beta} +{\cal G}_3(Q^2)q^{\alpha}]\gamma_5. \label{pscl}
\end{eqnarray}

In order to get helicity amplitudes in the rest frame of $f_1$, we introduce the four unit four-vectors in the Minkowski space, {\it i. e.}, 
\begin{equation}
\epsilon^{(t)} = \frac{q}{Q} \equiv \frac{1}{Q}(p_0,\vec{0}_{\perp},p), ~~~~~
\epsilon^{(0)} \equiv \frac{1}{Q}(p,\vec{0}_{\perp},p_0), ~~~~~
\epsilon^{(\pm1)} \equiv \mp\frac{1}{\sqrt{2}}(0,1,\pm i,0).
\end{equation}
Here $p_0$ = $\sqrt{p^2+Q^2}$. Now we introduce the completeness relation
\begin{equation}
\sum_{\mu} \epsilon^{(\mu)\dagger}_{\alpha} {\hat g}_{\mu\nu}\epsilon^{(\nu)}_{\beta} = g_{\alpha\beta},
\label{cmpl}
\end{equation}
where ${\hat g}_{\mu\nu}$ = 1 for $\mu$ = $\nu$ = $t$, -1 for $\mu$ = $\nu$ = 0, $\pm1$ and 0 otherwise.
By inserting Eq. (\ref{cmpl}) into the first Eq. (\ref{avd}), we get
\begin{equation}
A_V = \sum_{\mu} {\hat h}_V(m; \mu, \lambda; {\vec p})  D {\hat \ell}_V(\mu; \sigma; {\vec p}). 
\end{equation}
Here
\begin{equation}
{\hat h}_V(m; \mu, \lambda; {\vec p}) = h_V^{\alpha}(k_{\alpha} k^{\beta}-g_{\alpha}^{\beta})\epsilon^{(\mu)\dagger}_{\beta}
\end{equation}
and
\begin{equation}
{\hat \ell}_V(\mu; \sigma; {\vec p}) = {\hat g}_{\mu\nu}\epsilon^{(\nu)}_{\alpha} \ell^{\alpha}_V, 
\end{equation}
$\sigma$ denoting the set of possible quantum numbers (helicities, {\it etc.}) that characterize the decay products of $V$. 

Proceeding analogously to the scalar case, the various amplitudes 
${\hat h}_V(m; \mu, \lambda; {\vec p})$ result in
\begin{eqnarray}
{\hat h}_V(m;t,\lambda;{\vec p}) &=& \frac{1}{\sqrt{2\pi}} a_{t,\lambda}(p) 
{\cal D}^{1/2*}_{m-\lambda}(\phi,\theta,0),\\
{\hat h}_V(m;0,\lambda;{\vec p}) &=& \frac{1}{\sqrt{2\pi}} a_{0,\lambda}(p) 
{\cal D}^{1/2*}_{m-\lambda}(\phi,\theta,0),\\
{\hat h}_V(m;\pm1,\pm;{\vec p}) &=& \frac{1}{\sqrt{2\pi}} a_{\pm1,\pm}(p) 
{\cal D}^{1/2*}_{m\pm1/2} (\phi,\theta,0),
\end{eqnarray}
with
\begin{eqnarray}
\frac{1}{\sqrt{2\pi}} a_{t,\lambda}(p) &=& -\xi\frac{F}{Q}[p_0({\cal F}_A-2\lambda{\cal G}_A)-p({\cal F}_B-2\lambda{\cal G}_B)], \label{eq1} \\ 
\frac{1}{\sqrt{2\pi}} a_{0,\lambda}(p) &=& -\frac{F}{Q}[p({\cal F}_A-2\lambda{\cal G}_A)-p_0({\cal F}_B-2\lambda{\cal G}_B)], \label{eq2} \\ 
\frac{1}{\sqrt{2\pi}} a_{\pm1,\pm}(p) &=& -\sqrt{2}F\{-\chi {\cal F}_1(Q^2)+p{\cal F}_2(Q^2)\pm[{\cal G}_1(Q^2)+p\chi({\cal G}_2(Q^2)]\} \label{eq3}  
\end{eqnarray}
being the helicity amplitudes. Here
\begin{eqnarray}
{\cal F}_A &=& {\cal F}_1(Q^2)+\chi p {\cal F}_2(Q^2)+p_0{\cal F}_3(Q^2), \\
{\cal G}_A &=& \chi{\cal G}_1(Q^2)- p {\cal G}_2(Q^2)-\chi p_0{\cal G}_3(Q^2),\\
{\cal F}_B &=& \chi{\cal F}_1(Q^2)+ p {\cal F}_3(Q^2), \\
{\cal G}_B &=& {\cal G}_1(Q^2)- \chi p {\cal G}_3(Q^2)
\end{eqnarray}
and
\begin{equation}
\xi= 1-\frac{Q^2}{M^2}. \label{csi}
\end{equation}

\subsection{Remarks}

~~~~ A) If the boson $B$ decays into two spin-1/2 fermions, $B$ $\to$  $f'_1 ~~ f'_2$, the structures of the factors $\ell_S$, Eq. (\ref{asd}) and $\ell_V^{\beta}$, first Eq. (\ref{avd}), are, respectively, quite similar to (\ref{haS}) and to (\ref{haV}).

B) If $Q^2 << M^2$, the propagator $D$ is well approximated by a constant, 
$-M^{-2}$; this occurs typically for the gauge bosons in weak decays of hadrons and amounts to the current-current approximation. 

C) In the case of an off-shell vector boson, we are faced with its non-spin-1 component $t$ and with the helicity amplitude $a_{t,\lambda}$, involved, {\it e. g.}, in the charged pion decay. Moreover, the parameter $\xi$, Eq. (\ref{csi}), is well approximated by 1 if $Q^2 << M^2$, as occurs typically in the semi-leptonic decays of known hadrons. On the contrary, if $V$ is on-shell, $\xi$ vanishes and $a_{t,\lambda}$ is zero. In the semi-leptonic decay of the top quark, where the $W$ may be either on- and off-shell, the use of the complete formulae, that we have proposed just above, is recommended\cite{dg}. 

D) If $V$ = $W$, the propagator $D^{\alpha\beta}$, second Eq. (\ref{avd}), is a consequence of the choice of the unitary gauge, which is mandatory in the case of an off-shell boson\cite{ko}. Moreover, gauge invariance demands ${\cal F}_3$ = ${\cal G}_3$ = 0.

E) The terms proportional to $\lambda$ in Eqs. (\ref{eq1}) and (\ref{eq2}) and the term within square brackets in Eq. (\ref{eq3}) 
are parity-odd, since they correspond to the term (\ref{pscl}).


\section{On-shell Case - Polarizations in $f_1 \to f_2 S$ Decays} 

\subsection{Spin Density Matrix of $f_2$}

The spin density matrix of $f_2$ in the rest frame of $f_1$ reads as 
\begin{equation}
\rho_{\lambda\lambda'}^{f_2}(\theta,\phi;p) = \frac{1}{{\cal N}_0}  \sum_{m m'} {\hat h}_S(m; \lambda; \vec{p}) \rho^{f_1}_{m m'} {\hat h}_S^*(m'; \lambda'; \vec{p}).  \label{dm}
\end{equation}
Here ${\hat h}_S$ is given by Eq. (\ref{ria}) and ${\cal N}_0$ is a normalization constant, to be determined in a moment; moreover, 
\begin{equation}
\rho^{f_1} = \frac{1}{2}(I+{\vec \sigma} \cdot {\vec P}^{f_1}) \label{dm1}
\end{equation}
is the density matrix of the initial fermion $f_1$ and ${\vec P}^{f_1}$ its polarization, $|{\vec P}^{f_1}| \leq 1$. 
Eqs. (\ref{ria}), (\ref{dm}) and (\ref{dm1}) imply
\begin{equation}
\rho^{f_2}_{\lambda \lambda'} = \frac{1}{2\pi{\cal N}_0} a_{\lambda} {\rho'}_{-\lambda -\lambda'} a^*_{\lambda'}, \label{f2dmat}
\end{equation}  
with
\begin{equation}
\rho' = {\cal D}^{1/2}(\Omega)^{\dagger}\rho^{f_1}{\cal D}^{1/2}(\Omega) = \frac{1}{2}
[I + {\cal D}^{1/2}(\Omega)^{\dagger}{\vec\sigma} \cdot \vec{P^{f_1}} {\cal D}^{1/2}(\Omega)]. \label{P-r2}
\end{equation}
Here $\Omega$ $\equiv$ ($\theta$, $\phi$) and ${\cal D}^{1/2}(\Omega)$ = ${\cal D}^{1/2}(\phi,\theta,0)$. 
The probability conservation requires that  
\begin{equation}
\int d cos\theta d\phi \sum_{\lambda}\rho_{\lambda\lambda}^{f_2}(\theta,\phi;p) = 1, \label{mrm}
\end{equation}
which fixes 
\begin{equation}
{\cal N}_0 = |a_+|^2+|a_-|^2.
\end{equation}
Now we introduce the helicity frame, defined by the following three mutually orthogonal unit vectors:
\begin{equation}
\hat{e}_L = \frac{\vec{p}}{p}, ~~~ \hat{e}_N = \frac{\hat{k}\times\hat{e}_L}{|\hat{k}\times\hat{e}_L |}, ~~~ \hat{e}_T = \hat{e}_N \times \hat{e}_L; \label{unvc}
\end{equation}
here $\hat{k}$ is a unit vector in the direction of the $z$-axis of the rest (canonical) frame of $f_1$. Then, setting 
\begin{equation}
P_i = \vec{P^{f_1}}\cdot \hat{e}_i, ~~~ i = T, N, L, \label{ipol}
\end{equation} 
we get
\begin{equation}
\rho' = I + \sigma_1 P_T + \sigma_2 P_N + \sigma_3 P_L \label{P-r2}
\end{equation}
and Eqs. (\ref{f2dmat}) and (\ref{P-r2}) yield
\begin{equation}
\rho^{f_2}_{\pm\pm} = \frac{1}{4\pi{\cal N}_0}|a_{\pm}|^2(1\pm P_L), ~~~ \rho^{f_2}_{\pm\mp} = \frac{1}{4\pi{\cal N}_0} a_{\pm}a_{\mp}^*(P_T\mp iP_N). \label{dmf2}
\end{equation}

In comparing the theoretical expressions with data, one has to integrate the density matrix over $Q_1$, $Q_2$ and $Q$, which uniquely determine $p$: 
\begin{equation}
Q_1 = \sqrt{p^2+ Q_2^2} + \sqrt{p^2+ Q^2}. \label{peq}
\end{equation}
The integrated density matrix reads as   
\begin{equation}
{\bar\rho}^{f_2} (\theta,\phi) = \int dQ_1 w_1(Q_1) \int dQ_2 w_2(Q_2) 
\int dQ w_S(Q) \rho^{f_2} (\theta,\phi; p). \label{dmint} 
\end{equation}
Here each weight function $w(Q)$ is proportional to the modulus squared of the propagator of the particle - see Eq. (\ref{prp})\footnote{The weight functions of the initial and final fermions are related to their propagators, respectively, in the production process and in the secondary decay.} - , normalized so that $\int_0^{\infty} dQ w(Q)=1$. It coincides with the relativistic Breit-Wigner distribution function, which amounts to $2Q \delta(Q^2-m^2)$ for a stable particle. The integration (\ref{dmint}) is non-trivial in principle, since $\rho^{f_2} (\theta,\phi;p)$ depends on $p$ through the decay amplitudes and through $\rho^{f_1}$. However, the latter dependence may be generally neglected\cite{chu}; moreover, if the widths of the particles involved in the decay are sufficiently narrow - as it usually happens for on-shell particles, especially in weak decays -, also the former dependence is very mild. A different case will be treated in Sects. 5 and 6.


\subsection{Differential Decay Width and Polarization of $f_1$}

It is convenient to introduce the normalized differential width of the decay (\ref{ffv}), which we denote shortly as 
\begin{equation}
I(\Omega) = \frac{1}{\Gamma}\frac{d^2\Gamma}{d cos\theta d\phi}. \label{iom}
\end{equation} 
It results
\begin{equation}
I(\Omega) = \sum_{\lambda} \rho^{f_2}_{\lambda\lambda} = \frac{1}{4\pi}(1+\alpha^0_{as}P_L), ~~~~~~ 
\alpha_{as}^0 = \frac{\Delta {\cal N}_0}{{\cal N}_0} \label{angd}
\end{equation} 
where 
\begin{equation}
\Delta {\cal N}_0 = |a_+|^2-|a_-|^2.
\end{equation}
If parity is conserved, the asymmetry parameter $\alpha^0_{as}$ - the so-called analyzing power of spin\cite{ab1} - vanishes and the polarization of the parent resonance cannot be determined by means of the observable (\ref{iom}). If the decay is weak, this polarization can be determined, either by fitting the first expression (\ref{angd}) to the experimental data, or by means of the asymmetry\cite{dsa} \begin{equation}
{\cal A} = \frac{N(p_e>0)-N(p_e<0)}{ N(p_e>0)+N(p_e<0)}, \label{asp}
\end{equation}
where $p_e$ = $\vec{p}\cdot\hat{e}$, $\hat{e}$ is a unit vector and $N(p_e>0)$ [$N(p_e<0)$] is the number of events for which $p_e$ is positive (negative). This observable is related to $\vec{P}^{f_1}$, as follows from Eq. (\ref{angd}): 
\begin{equation}
{\cal A} = \alpha^0_{as} \vec{P}^{f_1}\cdot\hat{e}.
\end{equation}

\subsection{Polarization of $f_2$}

This observable is defined as 
\begin{equation}
I(\Omega){\vec P}^{f_2}(\Omega) = \sum_{\lambda\lambda'}\rho^{f_2}_{\lambda\lambda'} 
{\vec \sigma}_{\lambda'\lambda}. \label{pol}
\end{equation} 
By considering the components of the polarization along the unit vectors (\ref{unvc}), and substituting Eq. (\ref{f2dmat}) into Eq. (\ref{pol}), we get

\begin{eqnarray}
I(\Omega) P^{f_2}_L(\Omega) &=& \frac{1}{4\pi}(P_L+\alpha^0_{as}), ~~~~~~ ~~~~~~ \label{polcpL}
\\
I(\Omega) P^{f_2}_T(\Omega) &=& \frac{|\Phi_0|}{2\pi}(cos\psi_0 P_T-sin\psi_0 P_N), \label{polcpT}
\\ 
I(\Omega) P^{f_2}_N(\Omega) &=& \frac{|\Phi_0|}{2\pi}(sin\psi_0 P_T+cos\psi_0 P_N), \label{polcpN}
\end{eqnarray}

where the $P^{f_2}_i$ ($i = T, N, L$) are defined analogously to Eq. (\ref{ipol}) and
\begin{equation}
\Phi_0 = |\Phi_0|exp(-i\psi_0) = \frac{1}{{\cal N}_0}a_+ a_-^*. \label{par}
\end{equation} 
It is interesting to consider the orthogonal polarization vector of $f_1$, ${\vec P}_{\perp}$ = $P_T\hat{e}_T+ P_N\hat{e}_N $, and the analogous one for $f_2$. Indeed, Eqs. (\ref{polcpT}) and (\ref{polcpN}) imply that ${\vec P}^{f_2}_{\perp}$ is rotated with respect to ${\vec P}_{\perp}$ by an angle equal to the phase $\psi_0$, Eq. (\ref{par}). In this connection, we stress that a kinematical feature is strictly related to a dynamical one, owing to the fact that different helicity states undergo different interactions, possibly producing different phases. Moreover, $|{\vec P}^{f_2}_{\perp}|$ $<$ $|{\vec P}^{f_1}_{\perp}|$, as results from Eqs. (\ref{polcpT}) to (\ref{par}) and the Schwartz inequality.
 
At this point, three remarks are in order.

A) The phase $\psi_0$ - which as seen corresponds to a geometrical angle - is a $T$-odd observable, as results from Eqs. (\ref{polcpT}) and (\ref{polcpN}). Indeed, if as usual one chooses $\hat k$ along a given momentum, $P_N^{f_2}$ is $T$-odd, $P_T^{f_2}$ is $T$-even and $\psi$ changes sign under $T$-inversion. But also a different convention, such that $P_N^{f_2}$ is $T$-even and $P_T^{f_2}$ $T$-odd, would give rise to the same result. Therefore, we conclude that the phase is intrinsically $T$-odd. 

B) If $f_1$ is unpolarized, the polarization of $f_2$ can be only longitudinal and is nonzero only if parity is violated. On the other hand, the orthogonal polarization ${\vec P}^{f_2}_{\perp}$ may be present also in strong and electromagnetic decays; in these cases, it can be used for revealing the polarization of the parent resonance, which, as shown in Subsect. 2.2, cannot be detected by the differential decay width.

C) Last, considering the normalized amplitudes $\bar{a}_{\lambda}$ = $ a_{\lambda}/\sqrt{{\cal N}_0}$, we observe that their moduli and relative phase can be determined from the coefficients that appear in the expressions of $I({\Omega})$ (or of $P_L^{f_2}$) and of $P_{T(N)}^{f_2}$, taking account of the relation $|\bar{a}_+|^2+|\bar{a}_-|^2=1$.

\section{On-shell Case - Polarizations in $f_1 \to f_2 V$ Decay}

We take account of the first Eq. (\ref{haV}) and proceed analogously to the previous section; then, the spin density matrix of the ($f_2$-$V$) system reads as
\begin{equation}
\rho^{\mu\mu'}_{\lambda \lambda'} = \frac{1}{2\pi{\cal N}_1}\delta_{\Lambda, \mu-\lambda} a_{\mu,\lambda} {\rho'}_{\Lambda, \Lambda'} \delta_{\Lambda', \mu'-\lambda'} a*_{\mu',\lambda'}, \label{vdmat}
\end{equation} 
where ${\rho'}$ is given by Eq. (\ref{P-r2}) and $a_{\mu,\lambda}$ are the helicity amplitudes introduced in Subsect. 2.3, with $\lambda$, $\lambda'$ = $\pm1/2$ and $\mu$, $\mu'$  = 0, $\pm1$. By assuming the same approximations discussed in Subsect. 3.1, we get 
\begin{equation}
I(\Omega) = \sum_{\mu,\lambda} \rho^{\mu\mu}_{\lambda \lambda} = \frac{1}{4\pi}(1+\alpha^1_{as}P_L), ~~~~~~ \alpha^1_{as} = 
\frac{\Delta{\cal N}_1}{{\cal N}_1} \label{iom2}
\end{equation}
where, in this case,
\begin{equation}
{\cal N}_1 = |a_{+1,+}|^2+|a_{-1,-}|^2+|a_{0,+}|^2+|a_{0,-}|^2,  ~~~  \Delta {\cal N}_1 = 
|a_{+1,+}|^2+|a_{0,-}|^2-|a_{0,+}|^2-|a_{-1,-}|^2. \label{coefv}
\end{equation}
Moreover, the polarization of $f_2$ reads as
\begin{equation}
I(\Omega){\vec P}^{f_2}(\Omega) = \sum_{\lambda\lambda'\mu}\rho^{\mu\mu}_{\lambda\lambda'} 
{\vec \sigma}_{\lambda'\lambda}.
\end{equation}
Then,
\begin{eqnarray}
I(\Omega) P^{f_2}_L(\Omega) &=& \frac{1}{4\pi} ({\cal H}+
{\cal K} P_L), ~~~~~~ ~~~~~~ \label{polvL}
\\
I(\Omega) P^{f_2}_T(\Omega) &=& \frac{|\Phi_1|}{2\pi}(cos\psi_1 P_T-sin\psi_1 P_N), \label{polvT}
\\ 
I(\Omega) P^{f_2}_N(\Omega) &=& \frac{|\Phi_1|}{2\pi}(sin\psi_1 P_T+cos\psi_1 P_N), \label{polvN}
\end{eqnarray}
where
\begin{eqnarray}
{\cal H}({\cal K}) &=& \frac{1}{{\cal N}_1}(|a_{+1,+}|^2\pm|a_{0,+}|^2-|a_{0,-}|^2
\mp|a_{-1,-}|^2),
\\ 
\Phi_1 &=& |\Phi_1|exp(-i\psi_1) = \frac{1}{{\cal N}_1}a_{0,+}a_{0,-}^*.  \label{cfv2}
\end{eqnarray}

The behavior of the polarization of $V$ is different. We have  
\begin{equation}
I(\Omega){\vec P}^V(\Omega) = \sum_{\lambda\mu\mu'}\rho^{\mu\mu'}_{\lambda\lambda} 
{\vec S}_{\mu'\mu},
\end{equation}
where ${\vec S} \equiv (S_1, S_2, S_3)$, the $S_i$ denoting the spin matrices\cite{bal} for vector objects. Then
\begin{eqnarray}
I(\Omega) P^V_L(\Omega) &=& \frac{1}{4\pi} ({\cal J}+{\cal L} P_L), ~~~~~~ \ ~~~~~~ \ ~~~~~~ \ ~~~~ \label{polv2L}
\\
I(\Omega) P^V_T(\Omega) &=& \frac{\sqrt{2}}{4\pi}
[|\Phi^+|(cos\psi^+ P_T - sin\psi^+ P_N)+|\Phi^-|(cos\psi^- P_T + sin\psi^- P_N)], \label{polv2T}
\\ 
I(\Omega) P^V_N(\Omega) &=& \frac{\sqrt{2}}{4\pi}
[|\Phi^+|(cos\psi^+ P_N + sin\psi^+ P_T)+|\Phi^-|(cos\psi^- P_N - sin\psi^- P_T)], \label{polv2N}
\end{eqnarray}
where
\begin{equation}
{\cal J}({\cal L}) = \frac{1}{{\cal N}_1}(|a_{+1,+}|^2\mp|a_{-1,-}|^2), ~~~ \Phi^{\pm} = 
|\Phi^{\pm}|exp(\mp i\psi^{\pm}) = \frac{1}{{\cal N}_1}a_{\pm1,\pm}  a_{0,\pm}^*. \label{parv}
\end{equation}
We may re-write Eqs. (\ref{polv2T}) and (\ref{polv2N}) as
\begin{equation}
I(\Omega) P^V_{\perp i} (\Omega) =  \frac{\sqrt{2}}{4\pi}[|\Phi^+|R_{ij}(\psi^+)+
|\Phi^-|R_{ij}(-\psi^-)] P_{\perp j}, ~~~ (i,j = T,N),
\end{equation}
where $R$ is a matrix which describes a rotation around $\vec{p}$. It is worth observing that, if $\psi^+$ = $\psi^-$ = 0, $\vec{P}^V_{\perp}$ is parallel to $\vec{P}_{\perp}$, any deviation being due to a non-trivial phase. 

The remarks A) and B), at the end of Subsect. 3.3, hold also in the vector case; moreover the experimental measurements of $I(\Omega)$ and of the polarizations of $f_2$ and $V$ allow to determine the moduli of the 4 normalized amplitudes ${\bar a}_{\mu,\lambda}$ = $a_{\mu,\lambda}/{\cal N}_1$ and their relative phases. 

\section{Off-Shell Bosons and $V$-$S$ Interference}

Here we treat a decay of the type (\ref{ffv}) where $B$ consists of a scalar and a vector boson: one or both of them are virtual, they interfere and manifest themselves through a common decay mode. As an example, we consider the typical semi-leptonic decay
\begin{equation}
f_1 \to f_2 ~~ (W, ~ H) \to f_2 ~ \ell ~ \nu_{\ell}, \label{sld2}
\end{equation}
where the charged leptons $\ell$ with all momenta and directions are included.
In this case, the $p$-dependence of the density matrix of the ($f_2$-$B$) system ($B$ = $V$ and $S$) is no longer trivial, one has to integrate over $Q_1$, $Q_2$ and $Q$. We have
\begin{equation}
\tilde{\rho}(\Omega) = \int dQ_1 w_1(Q_1)\int dQ_2 w_2(Q_2)\int dQ w_V(Q)\hat{\rho}(\Omega;p).
\label{dmintv} 
\end{equation}
Here $\hat{\rho}(\Omega;p)$ is obtained from Eq. (\ref{vdmat}) by substituting
\begin{equation} 
a_{0,\lambda} \to a_{0,\lambda} + a_{t,\lambda} + x a_{\lambda}, \ ~~~~~~ \  {\cal N}_1 \to {\cal N}_v,
\label{vampt}
\end{equation}
where 	 
\begin{equation} 
x = B_S(Q)/B_V(Q), \ ~~~~~~ \ {\cal N}_v = |a_{+1,+}|^2+|a_{-1,-}|^2+
\sum_{\lambda}|a_{0,\lambda} + a_{t,\lambda} + x a_{\lambda}|^2 \label{symb}
\end{equation}
and, according to Sect. 2, $B_{V(S)}$ is proportional to the propagator of the vector (scalar) boson, normalized in such a way that $|B_{V(S)}|^2$ = $w_{V(S)}$. The expressions (\ref{iom2}),
for the differential decay width, and (\ref{polvL})-(\ref{polvN}) and (\ref{polv2L})-(\ref{polv2N}), for the components of the polarizations of the final particles, must be modified according to the substitutions (\ref{vampt}) and to the integration (\ref{dmintv}); however, such transformations are linear, therefore the structure of those expressions is preserved.

The present treatment generalizes the cases of Sects. 3 and 4. The integration (\ref{dmintv}) is generally trivial as regards the variables $Q_1$ and $Q_2$, since the particles $f_1$ and $f_2$ are on-shell and usually have narrow widths. Moreover, the integration over $Q$ does not affect the polarization of $f_1$\cite{chu}.

\section{Measurements of Polarizations of Final Particles}

The polarization of $f_2$ can be determined by means of a (weak) secondary decay of the fermion, analogously to the method described in Subsect. 3.2. In the case of the semi-leptonic top quark decay, first Eq. (\ref{sld}), the orthogonal polarization of the $b$-quark could be measured by exploiting its relation to the azimuthal dependence of the momentum of a hadron (say the $B$ meson) produced in the quark fragmentation\cite{col}.

As regards the polarization of the boson $V$, one has to elaborate a more complex strategy, especially if it is virtual and, possibly, interferes with a scalar boson $S$. As an example, consider the case where these bosons decay into two spin-1/2 fermions: 
\begin{equation}
f_1\to f_2 ~~~ (V, S) \to f_2 ~~~ f'_1 ~ f'_2, \label{dec2}
\end{equation}
which includes the semi-leptonic decay of the type (\ref{sld2}). 

We define a frame - to be denoted by ${\cal F}$ - at rest with respect to the bosons; here the momentum of $f'_1$ reads as
\begin{equation}
\vec{q} \equiv (q ~ sin \theta' ~ cos \phi', ~~ q ~ sin \theta' ~ sin \phi', ~~ q ~ cos \theta'),
\end{equation}
where $Q$ =$ \sqrt{q^2+m_1^2}+\sqrt{q^2+m_2^2}$ and $m_i$  is the mass of $f'_i$ ($i$ = 1, 2).
Then, we consider the joint distribution $I(\Omega,\Omega')$, with $\Omega'$ $\equiv$ ($\theta'$, $\phi'$). This is normalized in such a way that
\begin{equation}
\int d\Omega \int d\Omega' ~ I(\Omega,\Omega')=1,
\end{equation}
with $d\Omega$ = $d cos\theta d\phi$ and $d\Omega'$ defined analogously. Using the formalism introduced in the previous sections, we get
\begin{equation}
I(\Omega,\Omega')= \int d{\cal G}\sum_{\lambda'_1,\lambda'_2,\mu} [\rho'^B(\Omega,\Omega';p)]^{\mu\mu} \beta_{\lambda'_1,\lambda'_2}(q)\delta_{\lambda'_1-\lambda'_2,\mu}. \label{iid}
\end{equation} 
Here $\lambda'_i$  is the helicity of the fermion $f'_i$,  
\begin{equation}
\beta_{\lambda'_1,\lambda'_2}(q) = \frac{1}{\cal N'}|b_{\lambda'_1,\lambda'_2}(q)|^2, ~~~  
{\cal N'} = \sum_{\lambda'_1,\lambda'_2}|b_{\lambda'_1,\lambda'_2}(q)|^2
\end{equation}
and $b_{\lambda'_1,\lambda'_2}(q)$ are the helicity decay amplitudes for the secondary decay of the bosons\footnote{We have $b_{\lambda'_1,\lambda'_2}$ = $b^V_{\lambda'_1,\lambda'_2}+ b^S_{\lambda'_1,\lambda'_2}$, where $b^S_{\lambda'_1,\lambda'_2}$ is zero for the combinations $\lambda'_1 =  \lambda'_2 = \pm$.}. Moreover,
\begin{equation}
\rho'^B (\Omega,\Omega';p) = {\cal D}^1(\Omega')^{\dagger}  \rho^B(\Omega;p) {\cal D}^1(\Omega'), \label{rhp}
\end{equation} 
and
\begin{equation}
\rho^{B\mu\mu'}(\Omega;p)=\sum_{\lambda}{\hat\rho}^{\mu\mu'}_{\lambda \lambda}(\Omega;p) \label{dmsec}
\end{equation}
is the density matrix of the bosons, with ${\hat\rho}(\Omega;p)$ being defined in the previous section and 
${\cal D}^1$ being the Wigner rotation matrix for spin-1 objects. Last, $d{\cal G}$ denotes synthetically the integration (\ref{dmintv}). Note that also the parameters $\beta_{\lambda'_1,\lambda'_2}$ depend on $Q$ through the momentum $q$. The density matrix $\rho^B(\Omega;p)$ may be parametrized as\cite{ab2} 
\begin{equation}
\rho^B(\Omega;p) = \frac{1}{12\pi} (I + \frac{3}{2}\vec{S}\cdot\vec{P}^B + \sqrt{\frac{3}{2}}
\sum_{i=1}^3 T_it_i). \label{pard1}
\end{equation}
Here 
\begin{equation}
T_i = S_3S_i+ S_iS_3, ~~~ i=1,2, ~~~~~~ T_3 = 3S_3^2 - 2I, \label{tns}
\end{equation}
where $P^B_i$ and $t_i$ ($i$ = 1,2,3) are functions of $\Omega$ and $p$. More precisely, $P^B_i$ are the components according to the frame ${\cal F}$ of the polarization vector of $B$ and $t_i$ are 3 of the 5 components, according to the same frame, of the  irreducible alignment tensor\cite{ab2}. 

Inserting the parametrization (\ref{pard1}) and Eq. (\ref{rhp}) into Eq. (\ref{iid}) yields
\begin{equation}
I(\Omega,\Omega') = \frac{1}{16\pi^2}[1 + \alpha^B_{as} {\bar P}^B_L + \alpha_t {\bar{\cal T}}], \label{ii2}
\end{equation}
where
\begin{eqnarray}
\alpha^B_{as} {\bar P}^B_L &=& \frac{3}{2}\int d{\cal G} (|\beta_{++}|^2-|\beta_{--}|^2) P^B_L, 
~~~~~~ \ ~~~~~~ \ \label{pbvt}
\\
\alpha_t {\bar{\cal T}}&=& \sqrt{\frac{3}{2}}\int d{\cal G} [|\beta_{++}|^2+|\beta_{--}|^2- 
2(|\beta_{+-}|^2+|\beta_{-+}|^2)] {\cal T}, \label{tnsp}
\\
{\cal T} &=& z_L t_L -\frac{1}{2} t_3 (1-z_L^2). ~~~~~~ \ ~~~~~~ \ ~~~~~~ \ ~~~~~~ \ ~~~~~~ \ ~~~~~~ \ ~~~~~~ \ 
\label{calt}
\end{eqnarray}
Here, ${\bar P}^B_L$ and$z_L$ denote the projections along $\vec{q}$, respectively, of the average polarization and of $\hat{z}$, this latter being the unit vector along the $z$-axis of the frame ${\cal F}$. Furthermore, $t_L$ = $\sum_{i=1}^3 t_i q_i/q$, where $q_i$ are the components of $\vec{q}$ according to that frame. It is worth remarking that the rather complicated and apparently unnatural expression of ${\cal T}$, Eq. (\ref{calt}), corresponds to the fact that the three operators (\ref{tns}) are not the components of a vector, but rather of the alignment tensor\cite{ab2}.

The polarization of $B$ can be extracted from the distribution (\ref{ii2}) by two different methods:

a) by performing a best fit, using $\alpha^B_{as}{\bar P}^B_i$ and $\alpha_t {\bar t_i}$  ($i$ = 1, 2, 3) as free parameters\footnote{the bars indicate weighted averages, according to Eqs. (\ref{pbvt}) and (\ref{tnsp}).}: this may work if a wealth of data are available, typically with the top quark;
 
b) otherwise, by measuring, analogously to (\ref{asp}), the asymmetry, 
\begin{equation}
{\cal A}_B = \frac{N(q_r>0)-N(q_r<0)}{ N(q_r>0)+N(q_r<0)}, \label{aspv}
\end{equation}
with $q_r$ = ${\vec q}\cdot{\hat r}$ and ${\hat r}$ being a unit vector. Indeed, Eq. (\ref{ii2}) yields for the observable (\ref{aspv}) 
\begin{equation}
{\cal A}_B = \alpha^B_{as} {\vec P}^B\cdot \hat{r}. \label{asth}
\end{equation}

\section{Optimizing Data Analysis}

\subsection{Best Choice of the Frame}

As we have explained, an important observable for detecting signals of NP is the $T$-odd component of the polarization, which, according to our convention, coincides with $P_N^F$, where the suffix $F$ denotes either $f_2$ or the boson $V$, this latter possibly interfering with a scalar object. We are looking for the most convenient conditions for measuring it. In particular, if we choose  $\hat{k}$ $\propto$ $\vec{P}^{f_1}$, this observable turns out to be proportional to the $T$-odd triple product  $\vec{P}^F \cdot {\vec p} \times \vec{P}^{f_1}$, 
which consists only of quantities inherent in the decay (\ref{ffv}). Of course, this choice of the frame constitutes a drawback from the experimental point of view, as the direction of the polarization of $f_1$ is not known {\it a priori} and has to be determined by means of Eq. (\ref{angd}); however, it presents some important advantages, as we are going to illustrate. As a consequence of this choice, one has to set $P_N$ = 0 in Eqs. (\ref{polcpT})-(\ref{polcpN}), (\ref{polvT})-(\ref{polvN}) and (\ref{polv2T})-(\ref{polv2N}). 
This implies that $P_N^F$ is independent of the azimuthal angle and it is zero (maximal) when 
${\vec p}$ is parallel (orthogonal) to ${\vec P}^{f_1}$.

\subsection{Average Polarization}

In the frame that we have just defined, we choose an azimuthal plane passing through ${\vec P}^{f_1}$, where we fix the unit vectors $\hat{i}$ and $\hat{j}$ in such a way that $\hat{k}=\hat{i}\times\hat{j}$.
Then, in the half-space $0<\phi<\pi$, the average polarization reads as
\begin{equation}
\frac{1}{2}\vec{\cal P}^F = \int_0^{\pi} d\phi \int_0^{\pi} d\theta  sin \theta ~ I(\Omega) \sum_l P^F_l(\Omega){\hat e}_l, ~~~~~~~~~ l = T, N, L.
\label{avn}
\end{equation}

Considering separately the contributions corresponding to each unit vector ${\hat e}_l$, the results are  
\begin{eqnarray}
\vec{\cal P}^{f_2}_L &=& \frac{1}{2}\alpha_{as}^0\hat{j}+\frac{1}{3}P\hat{k} ~
(\frac{1}{2}{\cal H}\hat{j}+\frac{1}{3}{\cal K}P\hat{k}), ~~~~
\vec{\cal P}^B_L = \frac{1}{2}{\cal I}\hat{j}+\frac{1}{3}{\cal L}P\hat{k};\label{pol0}
\\
\vec{\cal P}^{f_2}_T &=& -\frac{4}{3}|\Phi_{0(1)}|cos\psi_{0(1)} P\hat{k},\ ~~~~~~ \ ~~~~~~ \ ~~~~~~
\vec{\cal P}^B_T = -\frac{2\sqrt{2}}{3}{\tilde\Phi}_T P\hat{k}; \label{pol1}
\\ 
\vec{\cal P}^{f_2}_N &=& -|\Phi_{0(1)}|sin\psi_{0(1)} P\hat{i},\ ~~~~~~ \ ~~~~~~ \ ~~~~ \ ~~~~~~
\vec{\cal P}^B_N = -\frac{1}{\sqrt{2}}{\tilde\Phi}_N P\hat{i}. \label{pol5}
\end{eqnarray}
Here, in the left column of Eqs. (\ref{pol0})-(\ref{pol5}), the expressions outside (inside) the parentheses refer to the $f_1$ $\to$ $S$($V$) $f_2$ decay; moreover $P$ = $|\vec{P}^{f_1}|$ and 
\begin{equation}
{\tilde\Phi}_T = |\Phi^+|cos\psi^+ + |\Phi^-|cos\psi^-, ~~~~~~~~~ \ 
{\tilde\Phi}_N = |\Phi^+| sin\psi^+ - |\Phi^-| sin\psi^-. \label{tn2}
\end{equation}

In the opposite half-space, ${\vec{\cal P}}^F_T$ and the component of ${\vec{\cal P}}^F_L$ along $\hat{k}$ are unchanged, while ${\vec{\cal P}}^F_N$ and the  component along $\hat{j}$ of ${\vec{\cal P}}^F_L$ change sign. Since $\hat{k}$ is taken along ${\vec P}^{f_1}$, Eqs. (\ref{pol1}) and (\ref{pol5}) show that, also after integration over the half-space, the component of ${\vec{\cal P}}^F$ in the azimuthal plane is rotated by some angle with respect to the polarization of the parent resonance. Obviously, in the case considered in Sect. 5, one has to modify the right column of Eqs. (\ref{pol0})-(\ref{pol5}) according to Eqs. (\ref{dmintv}) and (\ref{vampt}).   

\section{Applications}

~~~~ A) First of all, we consider semi-leptonic decays of the type (\ref{sld}), with polarized initial particles. If $T$-odd observables like $P_N^F$, or the phase $\psi_{0(1)}$, or the angle between $\vec{P}_{\perp}^B$ and $\vec{P}_{\perp}^{f_1}$ are significantly different than zero\cite{ga}, or than predictions of electromagnetic final state interactions\cite{iv2}, this is due to NP. In particular, the NP amplitude may be due to a charged Higgs contribution, or to a right-handed current, or to a tensor term, with a non-trivial phase with respect to the SM amplitude. Only a left-handed NP term would not give rise to any $T$-odd effect. In this connection, we observe that, if the decay $\Lambda_b \to \Lambda_c \tau \bar{\nu}_{\tau}$ confirms the anomaly found in the semi-leptonic decay of $B$\cite{bb1,bb2,bel}, a zero value for any of the above-mentioned $T$-odd observables would be an indication in favor of a $V-A$ NP contribution\cite{dsa1}.  

B) Such observables may imply also TRV; unlike other measurements which were suggested\cite{dug} few years ago, this implication is independent of the $CPT$ theorem. This last can be tested if data of the $CP$-conjugated semi-leptonic decays (say, those of $\bar{t}$ or $\bar{\Lambda_b}$) are available. 

C) Apart from NP, also the $T$-odd effects which are produced by the strong interactions\cite{ajd} are of some interest, as they are not negligible and, in general, not calculable analytically. These 
      effects can be determined by comparing the normal components of the polarization in
      a semi-leptonic and in a non-leptonic decay mode of the type 
      (\ref{ffv}), with $f_1$  and $f_2$ fixed; for example, one could 
      consider the $\Lambda_b$ decays (\ref{sld}) and (\ref{nld}). Also
      in this case NP could be revealed, provided one can compare such $T$-odd observables with those relative      
      to the $CP$-conjugated decay.
 
D) Similar NP tests could be performed by exploiting the production process of $f_1$. Indeed, considering the triple product $\vec{p}_{f_1}\times \vec{p}_b\cdot\vec{P}_{f_1}$, where $\vec{p}_{f_1}$ and $\vec{p}_b$ are the momenta, respectively, of $f_1$ and of the initial beam in the laboratory frame, we can define an asymmetry which is analogous to Eq. (\ref{asp}). If this asymmetry is significantly non-vanishing, and if it can be compared with the $CP$-conjugated one, the difference - if any - is due to TRV and, possibly, to NP. This kind of test is particularly indicated for the top and the anti-top quarks, which, under some conditions, are strongly polarized\cite{av,gmv,guv}.  

\section{Conclusions}

Here we resume the main results of our paper, concerning the (quasi-)two-body decays (\ref{ffv}). 

- We have derived the expressions of the differential decay width and of the polarization of the boson $B$ and of the fermion $f_2$, taking account of the widths of the various resonances. We have found that, if the initial fermion is polarized, the orthogonal component of the polarization of the final (spinning) particles is rotated, with respect to the same component of the polarization of $f_1$, by an angle which is related to the $T$-odd phase induced by the interactions involved in the decay. This generalizes a result that we found years ago\cite{dsa}.

- We have shown that, in the cases of spin-1/2 and spin-1 resonances, any component of the polarization amounts to an asymmetry. Incidentally, the polarizations of the initial and final fermion constitute useful sources of information, often neglected in the literature. 

- We have suggested a method for determining the $T$-odd observables by optimizing the data analysis. Such observables appear especially suitable for detecting NP, like the longitudinal polarization\cite{age} and the differential decay distribution\cite{jek} of the top quark in some kinematic regions. Also $T$-odd effects of strong interactions could be detected.
  
\vskip 0.25in

\centerline{\bf Acknowledgments}
The authors are thankful to their friend Flavio Fontanelli for useful and stimulating discussions.


\vskip 1cm
 
\begin{thebibliography}{0}
\bibitem{cms} S. Chatrchyan {\it et al.}, CMS Coll.: {\it Phys. Lett. B} {\bf 716} (2012) 30
\bibitem{atl} G. Aad {\it et al.}, ATLAS Coll.: {\it Phys. Lett. B} {\bf 716} (2012) 1
\bibitem{el} A. De Roeck {\it et al.}, {\it Eur. Phys. J. C}  {\bf 66} (2010) 525
\bibitem{wl} F. Wilczek, Proceedings {\it From the PS to the LHC Symposium on 50 Years of Nobel Memories in High-Energy Physics}, CERN,
Geneva, Switzerland, 3-4 Dec 2009, p. 189 
\bibitem{pe} M.E. Peskin: {\it Annalen der Phys.} {\bf 528} (2016) 20
\bibitem{alt} G. Altarelli: {\it Phys. Scripta T} {\bf 158} (2013) 014011
\bibitem{bb1} J.P. Lees {\it et al.}, BaBar Coll.: {\it Phys. Rev. Lett.} {\bf 109} (2012) 101802
\bibitem{bb2} J.P. Lees {\it et al.}, BaBar Coll.: {\it Phys. Rev. D} {\bf 88} (2013) 072012
\bibitem{bel} M. Huschle {\it et al.}, Belle Coll.: {\it Phys. Rev. D} {\bf 92} (2015) 072014
\bibitem{lhcb1} R. Aaij {\it et al.}, LHCb Coll.: {\it Phys. Rev. Lett.} {\bf 111} (2013) 191801
\bibitem{lhcb2} R. Aaij {\it et al.}, LHCb Coll.: {\it Phys. Rev. Lett.} {\bf 113} (2014) 151601
\bibitem{op} G. Abbiendi {\it et al.}, OPAL Coll.: {\it Phys. Lett. B} {\bf 444} (1998) 539
\bibitem{de} P. Abreu {\it et al.}, DELPHI Coll.: {\it Phys. Lett. B} {\bf 474} (2000) 205
\bibitem{atl2} M. Aaboud {\it et al.}, ATLAS Coll.: {\it JHEP} {\bf 1704} (2017) 124
\bibitem{gu1} T. Gutsche {\it et al.}: {\it Phys. Rev. D} {\bf 91} (2015) 074001; Erratum: 119907
\bibitem{ka} G.L. Kane {\it et al.}: {\it Phys. Rev. D} {\bf 45} (1992) 124
\bibitem{dg} R.H. Dalitz and G.R. Goldstein: {\it Phys. Rev. D} {\bf 45} (1992) 1531
\bibitem{bnr} W. Bernreuther {\it et al.}: {\it Nucl. Phys. B} {\bf 388} (1992) 53
\bibitem{jek} M. Jezabek and J.H. Kuehn: {\it Phys. Lett. B} {\bf 329} (1994) 317

\bibitem{fi1} M. Fischer {\it et al.}: {\it Phys. Lett. B} {\bf 451} (1999) 406	 
\bibitem{fi2} M. Fischer {\it et al.}: {\it Phys. Rev. D} {\bf 65} (2002) 054036
\bibitem{gr1} S. Groote {\it et al.}:  {\it Phys. Rev. D} {\bf 76} (2007) 014012
\bibitem{ab1} J.A. Aguilar-Saavedra and J. Bernabeu: {\it Nucl. Phys. B}{\bf 840} (2010) 349
\bibitem{ab2} J.A. Aguilar-Saavedra and J. Bernabeu: {\it Phys. Rev. D} {\bf 93} (2016) 011301
\bibitem{age} J.A. Aguilar-Saavedra {\it et al.}: {\it Phys. Lett. B} {\bf 769} (2017) 498
\bibitem{cz}  A. Czarnecki {\it et al.}: {\it Phys. Rev. D} {\bf 97} (2018) 094008
\bibitem{sh}  J. Shelton: {\it Phys. Rev. D} {\bf 79} (2009) 014032
\bibitem{gr2} S. Groote {\it et al.}: {\it Phys. Rev. D} {\bf 83} (2011) 054018
\bibitem{cho} D. Choudhury {\it et al.}: {\it Phys. Rev. D} {\bf 84} (2011) 014023
\bibitem{cao} J. Cao {\it et al.}:  {\it Phys. Rev. D} {\bf 85} (2012) 014025
\bibitem{kal} L. Kaldamae {\it et al.}: {\it Phys. Rev. D} {\bf 94} (2016) 114003

\bibitem{dug} G. Durieux and Y. Grossman: {\it Phys. Rev. D} {\bf 92} (2015) 076013
\bibitem{der} A. De Rujula {\it et al.}: {\it Nucl. Phys. B} {\bf 35} (1971) 365
\bibitem{ajd} Z.J. Ajaltouni and E. Di Salvo: {\it Int. Jou. Mod. Phys. E} {\bf 22} (2013) 1330006; 
 see also refs. therein
\bibitem{fi3} M. Fischer {\it et al.}: {\it Phys. Rev. D} {\bf 97} (2018) 093001


\bibitem{jk1} J.D. Jackson {\it et al.}: {\it Phys. Rev.} {\bf 106} (1957) 517
\bibitem{iv2} A.N. Ivanov {\it et al.}: {\it Phys. Rev. C} {\bf 98} (2018) 035503
\bibitem{br1} S.J. Brodsky {\it et al.}: {\it Phys. Lett. B} {\bf 530} (2002) 99
\bibitem{br2} S.J. Brodsky {\it et al.}: {\it Nucl. Phys. B} {\bf 642} (2002) 344
\bibitem{ag3} J.A. Aguilar-Saavedra and S.A. dos Santos: {\it Phys. Rev. D} {\bf 89} (2014) 114009
\bibitem{ga} R. Gatto: {\it Nucl. Phys.} {\bf 5} (1958) 183
\bibitem{av} O. Antipin and G. Valencia: {\it Phys. Rev. D} {\bf 79} (2009) 013013  
\bibitem{gmv} S.K. Gupta {\it et al.}: {\it Phys. Rev. D} {\bf 80} (2009) 034013
\bibitem{guv} S.K. Gupta and G. Valencia: {\it Phys. Rev. D} {\bf 81} (2010) 034013 

\bibitem{ag1} J.A. Aguilar-Saavedra: {\it Nucl. Phys. B} {\bf 812} (2009) 181
\bibitem{ag2} J.A. Aguilar-Saavedra {\it et al.}: {\it Eur. Phys. You. C}
{\bf 77} (2017) 200

\bibitem{dhg} L. Dhargyal: {\it Phys. Rev. D} {\bf 93} (2016) 115009
\bibitem{lee} J.-P. Lee: {\it Phys. Rev. D} {\bf 96} (2017) 055005
\bibitem{igt} S. Iguro and K. Tobe: {\it Nucl. Phys. B} {\bf 925} (2017) 560
\bibitem{chn} C.-H. Chen and T. Nomura: {\it Eur. Phys. Jou. C} {\bf 77} (2017) 631
\bibitem{val1} X.-G. He and G. Valencia: {\it Phys. Lett. B} {\bf 779} (2018) 52
\bibitem{val0} R. Martinez {\it et al.}: {\it Phys. Rev. D} {\bf 98} (2018) 115012

\bibitem{jw} M. Jacob and G.C. Wick: {\it Ann. Phys.} {\bf 7} (1959) 404
\bibitem{mi} L. Michel: Nuovo Cim. {\bf 14} Suppl. 1 (1959) 95

\bibitem{gu2} T. Gutsche {\it et al.}: {\it Phys. Rev. D} {\bf 88} (2013) 114018
\bibitem{gu3} T. Gutsche {\it et al.}: {\it Phys. Rev. D} {\bf 98} (2018) 053003
\bibitem{gu4} T. Gutsche {\it et al.}: {\it Phys. Rev. D} {\bf 98} (2018) 074011
\bibitem{gu5} T. Gutsche {\it et al.}: {\it Particles} {\bf 2} (2019) 339
\bibitem{ha} N. Habyl {\it et al.}: {\it Int. Jou. Mod. Phys. Conf. Ser.} {\bf 39} (2015) 1560112
\bibitem{ko} J.G. Koerner: {\it Proceedings, Helmholtz International Summer School on Physics of Heavy Quarks and Hadrons (HQ 2013)} JINR, Dubna, Russia, July 15-28 2013, p. 169; arXiv:1402.2787   

\bibitem{chu} S.U. Chung: {\it Phys. Rev.} {\bf 169} (1968) 1342
\bibitem{dsa} E. Di Salvo and Z.J. Ajaltouni: {\it Mod. Phys. Lett. A} {\bf 28} (2013) 1350048
\bibitem{bal} L.E. Ballentine: {\it Quantum Mechanics}. Prentice-Hall, 1990, p. 130
\bibitem{col} J.C. Collins: {\it Nucl. Phys. B} {\bf 396} (1993) 161
\bibitem{dsa0} E. Di Salvo and Z.J. Ajaltouni: {\it Mod. Phys. Lett. A} {\bf 24} (2009) 109
\bibitem{dsa1} E. Di Salvo, F. Fontanelli and Z.J. Ajaltouni: {\it Int. Jou. Mod. Phys. A} {\bf 33} (2018) 1850169 
\bibitem{be} J. Bernabeu and F. Martinez-Vidal: {\it Rev. Mod. Phys.}
{\bf 87} (2015) 165

\end{thebibliography}
\end{document}